\documentstyle[psfig,12pt]{article}
\begin{document}

\newcommand{\nc}{\newcommand}


\nc{\pa}{\partial}

\nc{\dpar}[2]{\frac{\partial{#1}}
                           {\partial{#2}}}
\nc{\lsim}
           {\begin{array}{c}\,\sim\hspace{2pt}\vspace{-19pt}\\< \end{array}}
\nc{\smlsim}
           {\begin{array}{c}\,\sim\hspace{2pt}\vspace{-18pt}\\< \end{array}}
\nc{\gsim}
           {\begin{array}{c}\,\sim\hspace{2pt}\vspace{-20pt}\\> \end{array}}
\nc{\un}{\underline}
\nc{\bs}{\boldsymbol}


\nc{\Do}{$\Downarrow$}
\nc{\ra}{\rightarrow} 
\nc{\Ra}{\Rightarrow}
\nc{\La}{Leftarrow} 
\nc{\lra}{\longrightarrow} 
\nc{\Lra}{\Longrightarrow} 
\nc{\lla}{\longleftarrow}
\nc{\Lla}{\Longleftarrow}  
\nc{\llra}{\longleftrightarrow} 
\nc{\Llra}{\Longleftrightarrow} 
\nc{\ora}{\overrightarrow}



\newcommand{\bsl}{$\backslash$}


\newcommand{\vsc}{\vspace{5mm}}
\newcommand{\vsd}{\vspace{10mm}}
\newcommand{\vsv}{\vspace{20mm}}
\nc{\mbh}{\mbox{\hspace{5mm}}}
\nc{\mbhd}{\mbox{\hspace{10mm}}}


\newtheorem{definition}{\bf Definition}[section]
\newtheorem{hypothesis}{\bf hypothesis}[section]
\newtheorem{principio}{\bf PRINCIPLE}[section]
\newtheorem{example}{\bf Example}[section]
\newtheorem{thesis}{\bf Thesis}[section]


\newfont{\pippo}{cminch}
\newfont{\adriana}{cmss17 scaled\magstep 5}

\bibliographystyle{unsrt}
\textwidth 12.5truecm
\textheight 19truecm

\begin{titlepage}
\begin{flushright}
INFN/TH-97/02
\end{flushright}
\vspace*{50mm}
\begin{flushleft}
{\bf ENTROPY PRODUCTION IN THE EARLY UNIVERSE} 
\vspace{3mm}

{\it A new mechanism}\\
     \vspace{6mm}          
     \hspace{2cm} P. DI BARI \\
     \hspace{2cm}{\it INFN-Laboratori Nazionali del Gran Sasso} \\
     \hspace{2cm}{\it Strada Statale 17 bis, 67010 Assergi (L'Aquila), Italy }
\end{flushleft}
\begin{abstract}
We expose briefly the role of entropy in the early universe and in particular 
the importance of searching for new mechanisms of entropy production. 
We describe a mechanism \cite{BeDi96} that shows how entropy is produced 
during early annihilations and under which conditions the production is not 
negligible. A MeV $\tau$ neutrino provides an interesting application.
  
\vspace{3mm}

This paper has been presented in the Nato School Session on ``Generation 
of Cosmological Large-Scale Structure" (Erice, November 1996) and it will
appear in the proceedings of the school (ed. D.N.Schramm).
\end{abstract}
\end{titlepage}

\setcounter{page}{2}

{\bf 1. Introduction}
\vspace{3mm}

Entropy is not a fundamental quantity as e.g. energy density, but
it provides an auxiliary function useful to describe the physical 
processes in the early universe. This is evident when thermodynamical 
equilibrium applies (quasi-static expansion) because entropy is conserved  and 
thus it represents a {\em constant of motion}. This is a traditional result 
\cite{We} that can be shown easily by applying to the cosmological 
expansion the usual thermodynamical relation $d(\rho R^{3})/dt=
TdS/dt-pdR^{3}/dt+\sum_{i}\mu _{i}dN_{i}/dt$
and imposing the energy-momentum tensor conservation equation
$d(\rho R^{3})/dt=-pdR^{3}/dt$. Then necessarily:
\begin{equation} \label{eq:dSdt}
\frac{dS}{dt}=-\frac{1}{T}\sum_{i} \mu_{i}\frac{dN_{i}}{dt}
\end{equation}
But as we are assuming the equilibrium also in the chemical reactions, 
the right sum is zero and hence $dS/dt=0$.
Entropy conservation allows for the straightforward derivation of the 
$R-T$ relation. 
If all species are in radiative equilibrium then $S\propto g_{R}(TR)^{3}$,
where $g_{R}=const$ are the {\em effective degrees of freedom}, and simply 
$TR=const$, otherwise, the definition of $g_{R}$ 
is generalized to the case in which some species are not ultrarelativistic 
by introducing a function $g_{S}(T)$ that can be easily calculated 
\cite{KT} and in this case 
$RT\propto [g_{S}(T)]^{-1/3}$. The $RT$ factor is important because
the particle numbers of any species in radiative equilibrium, and in
particular the photon number $N_{\gamma}$, is $\propto (RT)^{3}$.
Thus entropy conservation provides the way to know 
how the photon number changed during the early universe history.
During the annihilations of some particle species the degrees of freedom 
decrease and photons are produced because the disappearing species 
releases its entropy to the radiative plasma including photons. Sometimes
in the literature this is called entropy production while it is only 
entropy exchange. 
Photon production is an important quantity to be known because 
it is needed to calculate the relic abundances $(n/n_{\gamma})_{0}$ 
of particle species or of the barionic number. In fact if at some
time $t_{f}$ a particle or charge number $N_{f}$ is frozen, 
we are often able to calculate the abundance $(n/n_{\gamma})_{f}$
at that time and then the relative relic abundance can be expressed as
$(n/n_{\gamma})_{0}=(n/n_{\gamma})_{f}(N_{\gamma}^{f}/N_{\gamma}^{0})$.
This expression shows how photon production dilutes the abundance 
from freezing until the present. For this reason it is convenient to define
a {\em dilution factor} $f\equiv N_{\gamma}^{0}/N_{\gamma}^{f}$.
Now it is easy to understand the effect of an entropy production: it
gives a further contribution to photon production and hence to $f$:
$f= f_{g}\cdot f_{S}$ (where $f_{g}=g_{S}^{f}/g_{S0}$
and $f_{S}=S_{0}/S_{f}$).  
We point out that the more recent the entropy production is the
more effective it is, because it dilutes all densities previously frozen.

\vspace{6mm}
{\bf 2. Entropy production calculation through the L.W. equation}
\vspace{3mm}

Let us consider the classical problem of the freezing of particle abundances 
during annihilations. Some particle species $h$ can annihilate eventually
into photons through a lighter particle species
$l$: $h+\bar{h} \lra l+\bar{l} \lra \ldots \lra \gamma\mbox{'s}$.
 Assuming that the elastic reactions rates are strong enough to
support the kinetic equilibrium (thus a temperature is
defined for the $h$ and it is kept equal to photon temperature), 
the $h$ distribution function will assume an equilibrium form:
$f_{h}(\vec{p},t) \simeq \tilde{f}(\vec{p},t)=
\{e^{ \beta(t)[E_{\vec{p}}-\tilde{\mu}_{h}(t)]} + 1\}^{-1}$ 
where $\tilde{\mu}_{h}$ is commonly called the {\em pseudo-chemical} potential
to underline that it does not obey, in general, the chemical equilibrium 
condition $\mu_{h}+\mu_{\bar{h}}=0$. In this way the entropy production rate
can be obtained easily from thermodynamics using (\ref{eq:dSdt}), and in our 
specific case becomes:
\begin{equation}\label{eq:dSt}
\frac{dS}{dt}=-\frac{(\tilde{\mu}_{h}+\tilde{\mu}_{\bar{h}})}{T} 
               \frac{dN_{h}}{dt} 
\end{equation}  
This expression can be obtained from statistical mechanics as well. 
It can be noticed that in order to have
$dS/dt\neq 0$ two conditions must be satisfied: the first one is departure
from chemical equilibrium ($\tilde{\mu}_{h}+\tilde{\mu}_{\bar{h}}\neq 0$)
and the second is a decreasing particle number ($dN_{h}/dt\neq 0$) which means
the presence of annihilation processes not balanced by pair production
processes. 
If we want to perform a quantitative calculation of the entropy production, we
must use some model to calculate $\tilde{\mu}_{h}$ and $dN_{h}/dt$.
The simplest way to describe the annihilations is represented by the special
Lee-Weinberg equation:
\begin{equation}\label{eq:eLW}
\frac{d\bar{N}}{dy}=A_{0}\left[\bar{N}^{2}(y)-\bar{N}^{2}_{eq}(y)\right]
\end{equation}
($y=T/m$, $\bar{N}=N_{h}/N_{in}$, $N_{in}=N_{h}(T\gg m_{h})$,  
$A_{0}=0.055(g_{h}/\sqrt{g_{\rho}^{in}})\sigma_{0} m m_{PL}$).
In the modern literature this equation is usually written using 
the variable $Y=n/s$ ($s$ is the entropy density) where $Y\propto N$
if entropy is conserved. But as we are relaxing entropy conservation
assumption we must use a variable like $\bar{N}$. 
In this equation particle-antiparticle symmetry is assumed. This is not 
restrictive because it means only that the departure from equilibrium and
the consequent freezing must happen for temperatures when 
$(N_{h}-N_{\bar{h}})/N_{h}\ll 1$. Otherwise 
it is well known \cite{Ze65} that departure from equilibrium
would be greatly delayed and therefore entropy production would be negligible.
 Thus we can set 
$\tilde{\mu}_{h}=\tilde{\mu}_{\bar{h}}=\tilde{\mu}$ and the chemical
equilibrium condition becomes $\tilde{\mu}=0$. Moreover, it is 
an implicit approximation of the equation that 
$\tilde{z}\equiv \tilde{\mu}/m=y\ln[\bar{N}(y)/\bar{N}_{eq}(y)]$.
 In Fig. 1 we show the solutions $\bar{N}$ of Lee-Weinberg equation 
for different values of $A_{0}$ and the correspondent $\tilde{z}$ .
\vspace{-6mm}
\begin{figure}[hbt]
\hspace{5mm}
\begin{flushright}
\psfig{file=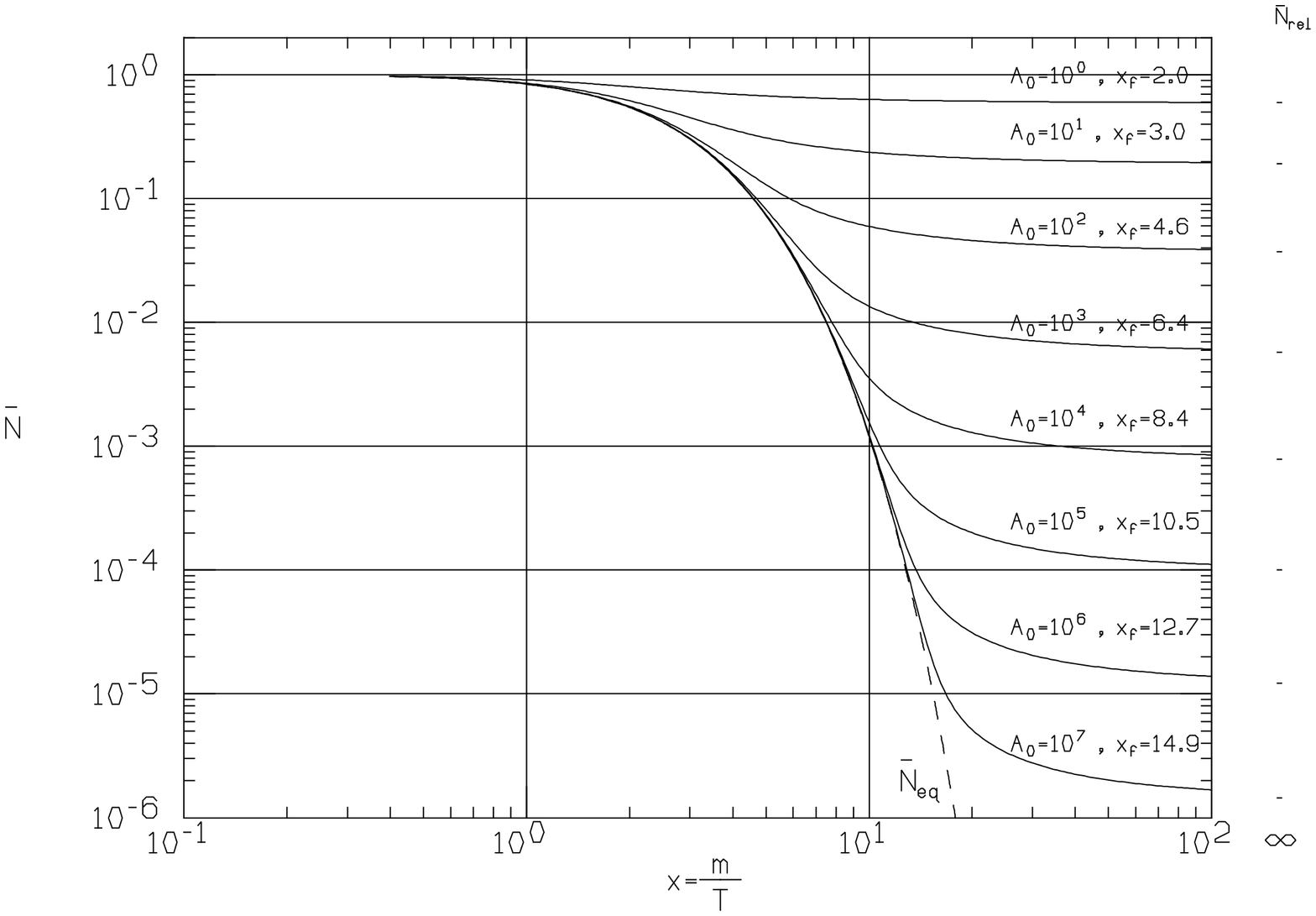,width=56mm,height=55mm,angle=90}
\vspace{-36mm}
\hspace{5mm}
\psfig{file=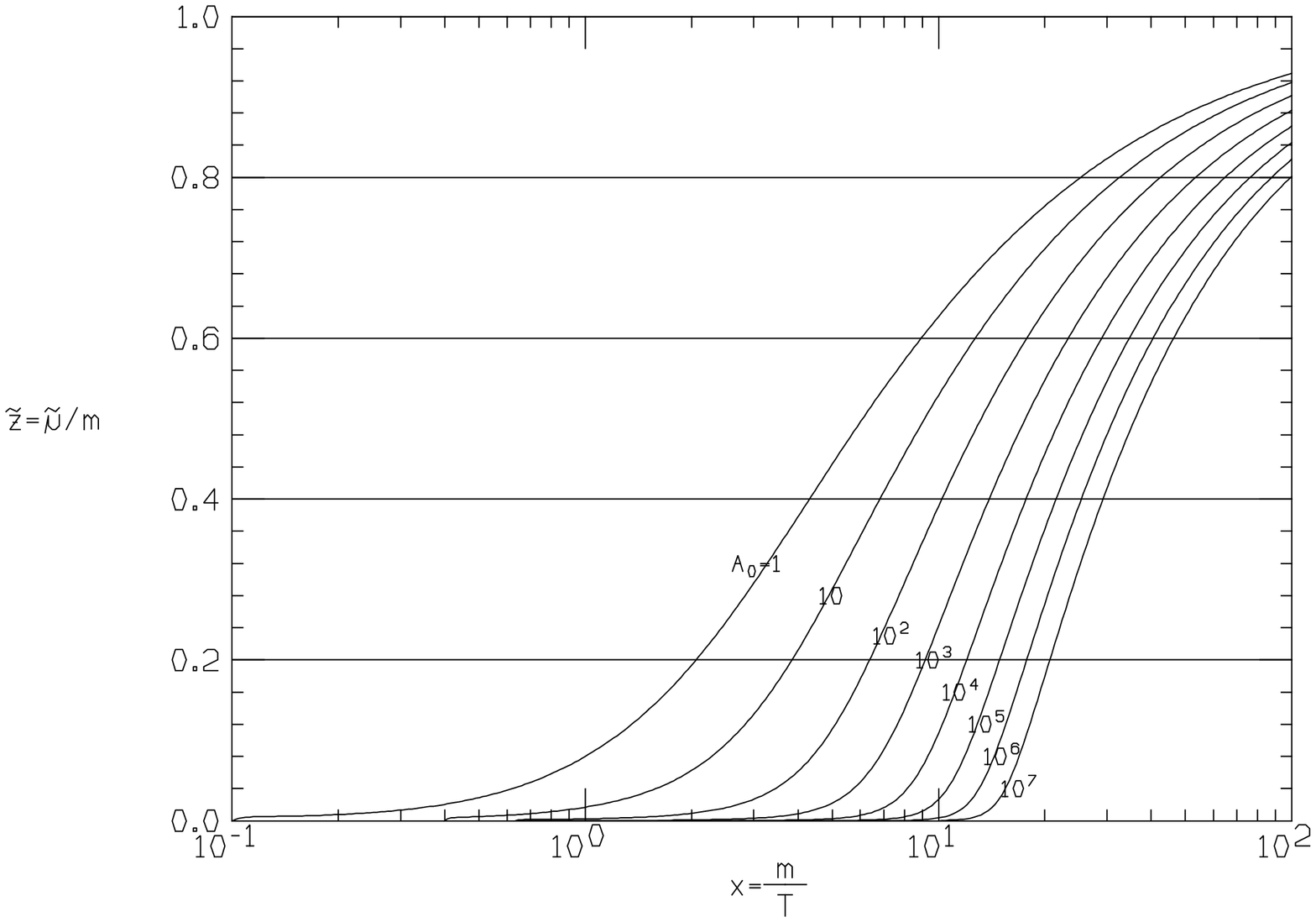,width=66mm,height=36mm,angle=90}
\end{flushright}
\end{figure}
\normalsize
\vspace{-3mm}
\begin{center}
{\small{\em Figure 1.}}
\end{center}

Finally we can write the entropy production rate through $\bar{N}$ obtaining:
\begin{equation}\label{eq:dSyS}
\frac{1}{S_{h}^{in}}\frac{dS(y)}{dy}=-\frac{135\xi(3)}{4\pi^{4}}
                              \frac{\tilde{z}(y)}{y}\frac{d\bar{N}(y)}{dy}
\end{equation}
where $S_{h}^{in}=g_{h}S_{in}/g_{S}^{in}$. By integration one can obtain
the entropy production $\Delta S(y,y_{in})$. The results are shown in Fig. 2.

The rates present a maximum because entropy production 
is zero before the system goes out of equilibrium (i.e. $\tilde{\mu}=0$) and 
after it tends again to zero because the annihilation rate decreases 
($dN/dt \ra 0$, while $\tilde{\mu}$ reaches its asymptotical value $m$). 
Moreover, one can notice the existence of a value for $A_{0}$ for which 
entropy production is maximum. In fact both for $A_{0}\ra 0$ and for 
$A_{0}\ra\infty$ entropy production also tends to zero: in the first case 
because particles would not annihilate at all, though the system is out of 
equilibrium, and in the second because the system never goes out of 
equilibrium. Thus there must exist a maximum.  
 It is reached when $A_{0}=3.65$ and $x_{f}=2.5$, where we defined $x_{f}$ 
as that value for which $\bar{N}-\bar{N}_{eq}=0.5 \bar{N}_{eq}$. The maximum 
production of entropy occurs for a freezing during which $h$-particles are 
semi-relativistic. 

\vspace{-6mm}
\begin{figure}[hbt]
\hspace{5mm}
\begin{flushright}
\psfig{file=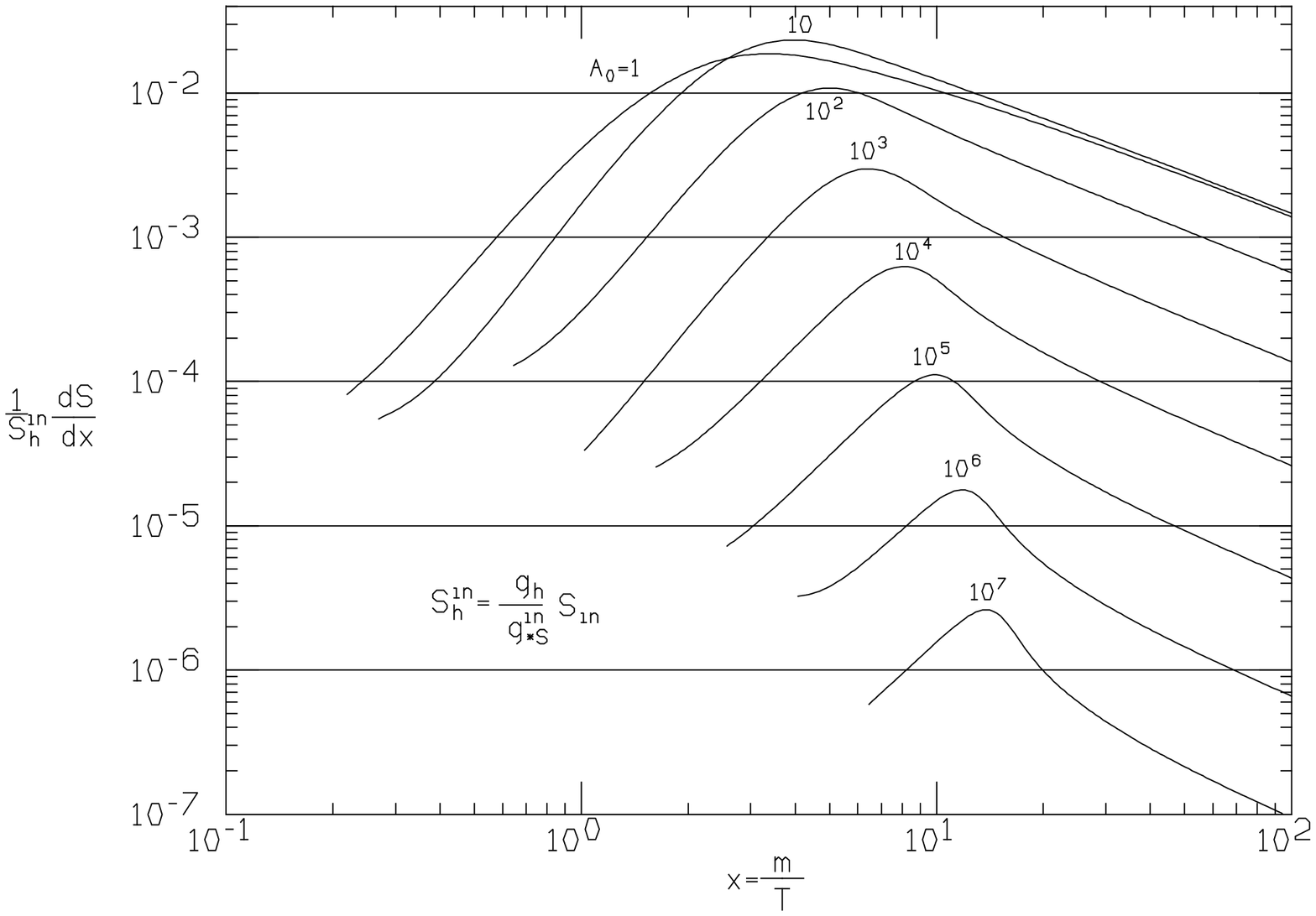,width=56mm,height=55mm,angle=90}
\vspace{-36mm}
\hspace{5mm}
\psfig{file=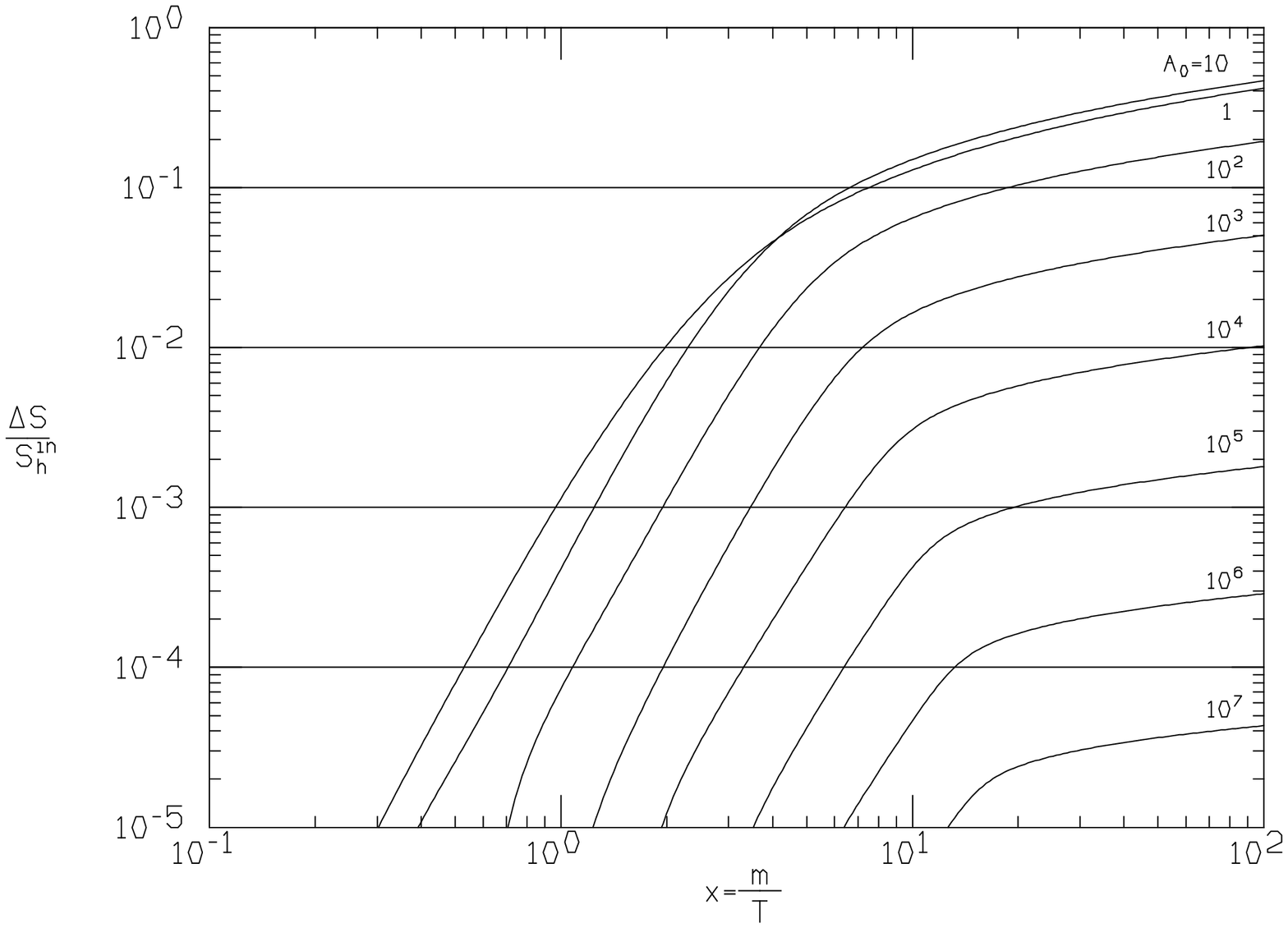,width=66mm,height=36mm,angle=90}
\end{flushright}
\end{figure}
\normalsize
\vspace{-3mm}
\begin{center}
{\small {\em Figure 2.}}
\end{center}

In the {\em asymptotical limit} ($x=m/T\ra\infty$) all expressions become 
simpler: $d\bar{N}(x)/dx \lra -A_{0}\bar{N}^{2}_{0}/x^{2}$, 
$\tilde{z}(x)\lra 1$ and $dS(x)/dx\ra \propto A_{0}\bar{N}^{2}_{0}/x$: hence
asymptotically the entropy production $\Delta S=\int dx (dS/dx)$ does not 
stop but it increases logarithmically, thanks to the relic annihilation. 

\vspace{6mm}
{\bf 3. Estimation of maximum entropy dilution factor}
\vspace{3mm}

In order to estimate the maximum dilution that the mechanism can produce
we must integrate the entropy production over the entire early universe
history, until matter-radiation decoupling.
Another consideration to be made is that during the early universe
the degrees of freedom decrease, and this requires the generalization 
of the calculation of $dN/dt$ done with the simple Lee-Weinberg equation 
which assumes degrees of freedom and entropy constant. 
At the same time we can also include the correction due to the
variation of entropy. 
It is possible to introduce these modifications only in the asymptotical 
regime (which gives the major contribution to the entropy production). 
Hence during this phase the decoupled equations (\ref{eq:eLW}) and 
(\ref{eq:dSyS}) can be replaced by two coupled equations:
\begin{equation}
\frac{d\bar{S}}{dy}=
-\frac{135}{4}\frac{\xi(3)}{\pi^{4}}\frac{g_{h}}{g_{S}^{in}}
\frac{1}{y}\frac{d\bar{N}}{dy}
\mbhd
\frac{d\bar{N}}{dy}=
A_{0} \cdot \frac{\bar{g}_{Si}}{\sqrt{\bar{g}_{\rho i}}}
\frac{1}{\bar{S}(y)}\bar{N}^{2}_{0}
\end{equation}
($\bar{S}=S/S_{in}$, $\bar{g}_{Si}=g_{Si}/g_{S}^{in}$, 
$\bar{g}_{\rho i}=g_{\rho i}/g_{\rho}^{in}$) that yield the solution:
\begin{equation}\label{eq:barS} 
\bar{S}_{dec}=\bar{S}_{1}=\bar{S}_{*}
\sqrt{1+2\frac{a}{\bar{S}^{2}_{*}}\frac{g_{h}}{g_{S}^{in}}\sum_{i=*,N,\dots,1}^{1}
\frac{\bar{g}_{Si}}{\sqrt{\bar{g}_{\rho i}}}\ln\frac{x_{i-1}}{x_{i}}} 
\mbh a=\frac{135}{4}\frac{\xi(3)}{\pi^{4}}A_{0}\bar{N}_{0}^{2}
\end{equation}

If the second term in the square root is much smaller than $1$ then, expanding 
at first order, the logarithmic behaviour  is restored.
Until now we supposed that all components are coupled to photons and 
therefore $f_{S}=\bar{S}_{dec}$. If there is a decoupled component it must be 
considered that some fraction of the entropy can be given to 
the decoupled component and thus does not contribute to the dilution. 

Now we must find the maximum value that the expression (\ref{eq:barS}) can
assume. It depends on two parameters: $A_{0}$ and $T_{in}$ ($g_{h}$ 
can be set equal $2$). We have already indicated that $A_{0}=3.65$ yields
the maximum  value for $A_{0}\bar{N}^{2}_{0}$. To find the appropriate $T_{in}$
there are two opposite considerations: to get the greatest temperature range 
requires annihilations as early as possible, but to have the greatest 
fractional weight of $h$-particles ($\equiv g_{h}/g_{S}^{in}$) 
requires annihilations that happen when as few particle species as possible 
have survived.
 The second consideration suggests to take a time 
interval such that $g_{S}^{in}$ is minimum, and the first 
suggests the insertion of annihilations as early as possible inside this 
interval.
 Thus the best situation is when freezing starts soon after 
electron-positron annihilations so that $m\lsim m_{e}/2 \simeq 0.25 MeV$
and from (\ref{eq:barS}) $f_{S}=1.43$ is obtained. The second possibility
is that freezing starts soon after muon annihilations
($m_{h}\lsim m_{\mu}/2 \simeq 50 MeV$) and in this case  
$f_{S}=1.31$ is obtained.

\vspace{6mm}
{\bf 4. Application: Mev $\tau$-neutrino}
\vspace{3mm}

In this section we want to consider a realistic case that produces a dilution 
factor as much as possible close to the order of magnitude found earlier. 
This time we must substitute $\sigma_{0}=const$ with the thermally averaged 
cross section $\langle\sigma_{ann}v_{Mo}\rangle (y)$, 
that can be calculated with a single-integral formula \cite{GoGe91} 
valid for any value of $x_{f}$. 
  Unfortunately it is not realistic an hundred Kev particle that 
undergo the semirelativistic
freezing ($x_{f}\simeq 2$) we need, though further investigations are in 
progress considering Majoron or Axino.
 An Mev $\tau$-neutrino is a particle that would freeze with the right
$x_{f}$ and it is light enough to have a good fractional weight.
 This then seems to be the best application of the mechanism we described and 
also because of the possible consequences that it could have on its mass 
constraints. In last six years constraints from nucleosynthesis 
\cite{Kolb91,Fields96}, SN1987A \cite{Burro92},  
have been greatly improved, reducing the possibilities for an MeV neutrino
from an astrophysical point of view.
At the same time the laboratory experiment lower limit has been 
lowered to $24$ MeV  \cite{ALEPH95} and thus in future years there could be 
an important test for astroparticle physics. This is why we think that it is 
worthwhile to continue investigations on MeV $\tau$ neutrino.
Here, however, we want mainly to provide an application for the general 
mechanism we described.

\vspace{-7mm}
\begin{figure}[hbt]
\hspace{5mm}
\begin{flushright}
\psfig{file=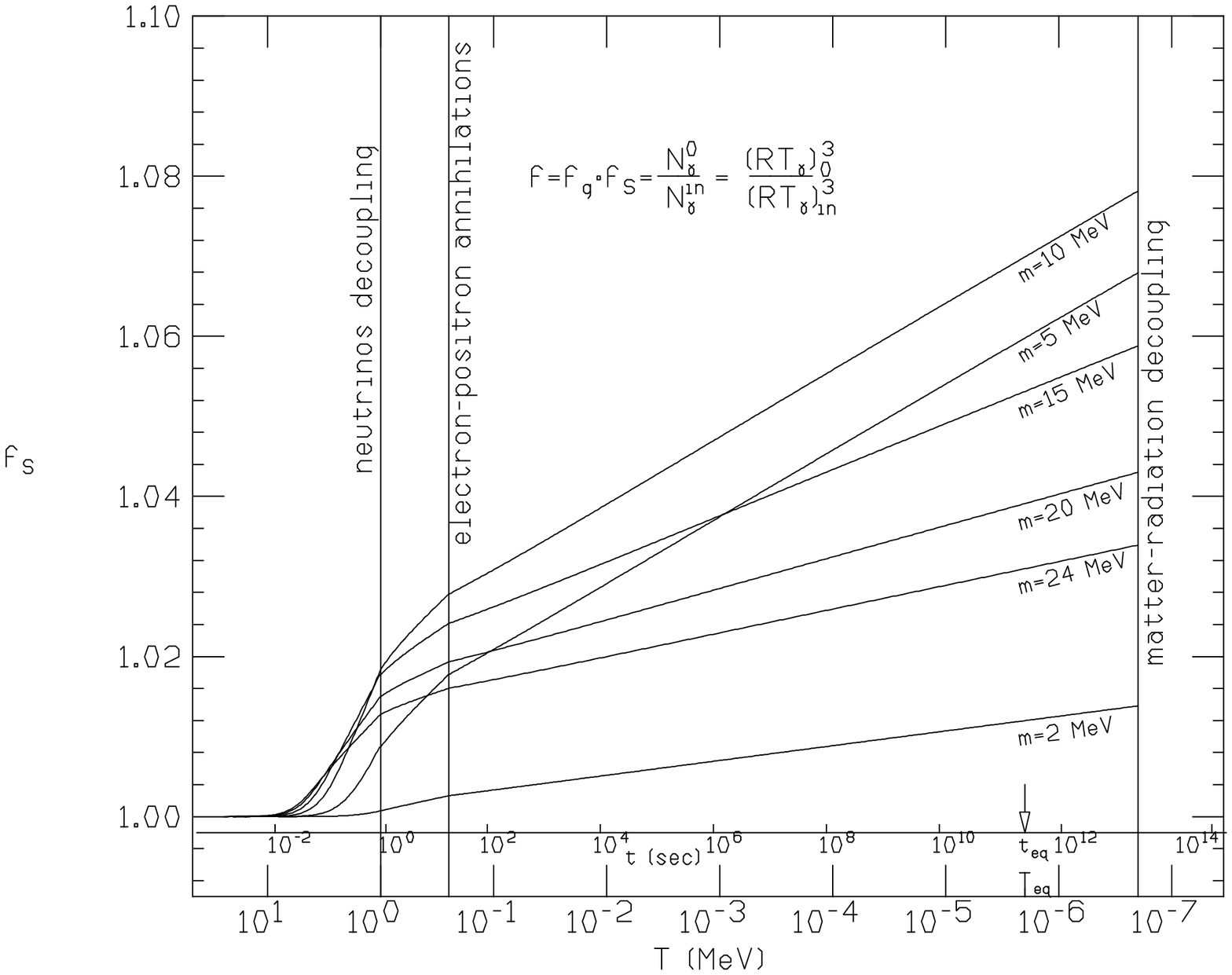,width=58mm,height=53mm,angle=90}
\vspace{-36mm}
\hspace{5mm}
\psfig{file=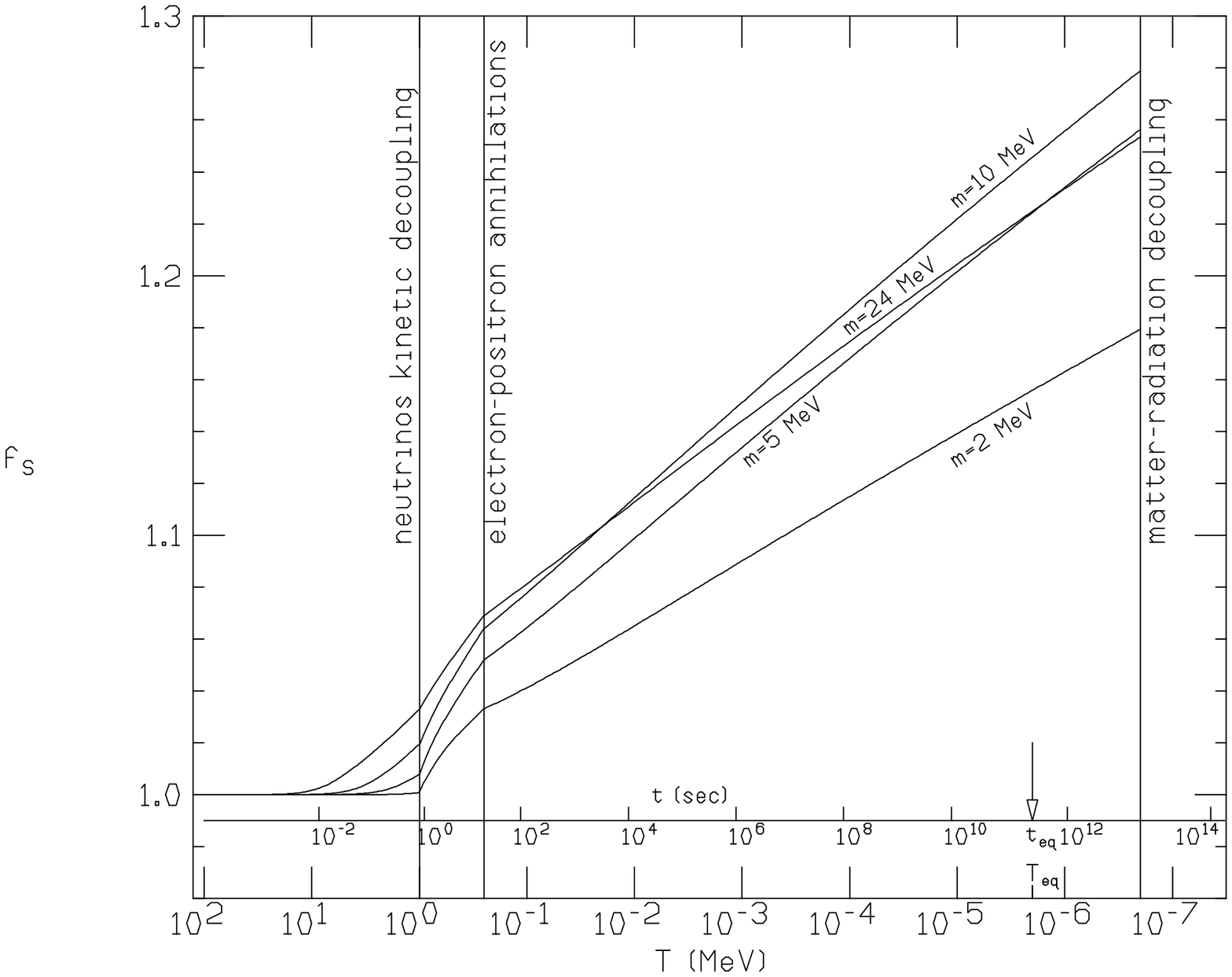,width=68mm,height=36mm,angle=90}
\end{flushright}
\end{figure}
\normalsize
\vspace{9mm}
\begin{center}
{\small {\em Figure 3.} Entropy dilution factor: $MeV$ Dirac 
$\nu_{\tau}$ (left); $MeV$ Giudice $\nu_{\tau}$ (right).}
\end{center}

 We performed the calculations for a Dirac neutrino and for 
a neutrino with a magnetic moment 
$\mu \gsim 2\cdot 10^{-9} (m_{\nu}/10MeV)\mu_{B}$ \cite{Giudice90}.
In the first case it must be considered that $\nu_{\tau}$ can annihilate
into $e^{+}e^{-}$, $\nu_{\mu}\bar{\nu}_{\mu}$, $\nu_{e}\bar{\nu}_{e}$ 
and that for $T\lsim 1 MeV$ the neutrinos are decoupled: this 
reduces of a factor $\simeq 1/3$ the dilution factor from the maximum
and the result is $f_{S}=1.01\div 1.08$ for masses $m=2\div 24 MeV$
(see Fig.3 left).
In the second case the annihilation through photon exchange 
in $e^{+}e^{-}$ is dominant. In this case the factor $1/3$ is avoided
and the dilution factor can approach the maximum value permitted 
($f_{S}=1.1\div 1.3$ for masses $m=2\div 24 MeV$ as shown in Fig.3 right).  

\vspace{6mm}
{\bf 5. Conclusions}
\vspace{3mm}

In the introduction we clarified the difference between the
simple entropy transfer from the annihilating particle species to the 
radiative plasma and an effective entropy production.
We presented here the possibility to have a significant entropy production
in the late phase of the early universe if the annihilating 
particle species undergoes a semirelativistic freezing.
 This mechanism can dilute all relic abundances in the early universe
by a factor $1.5$ as maximum value. In the last section we presented an 
application that realizes this possibility: an $MeV$ $\nu_{\tau}$.
 In the general cases of nonrelativistic and 
ultrarelativistic freezing, entropy production is a negligible 
effect as usually assumed.

\vsc

This talk is based on the work \cite{BeDi96}. 
P. Di Bari has been fortunate to have V.Berezinsky as the supervisor in his
``Tesi di Laurea''.  He wishes to thank A.D. Dolgov for valuable 
suggestions and M. Lusignoli for his helpful role during the work of thesis. 
He also thanks all people and the director of Gran Sasso Laboratories 
P. Monacelli for their warm hospitality during the last two years and Lori 
Gray for her help with the English.

\end{document}